# Full-field moisture induced deformation in Norway spruce: Intra-ring variation of transverse swelling

Christian Lanvermann, Falk K. Wittel, Peter Niemz

*ETH Zurich, Institute for Building Materials, Schafmattstrasse 6, 8093 Zurich, Switzerland*

Phone: +41 44 632 32 798; Fax: +41 44 632 11 74; lanvermannchr@ethz.ch

The transverse hygro-expansion of the Norway spruce wood is studied on the growth ring level using digital image correlation. This non-destructive technique offers the possibility to contactless study deformation fields of relatively large areas. The measured full-field strains are segmented into individual growth rings. Whereas radial strains closely follow the density progression with the maximum in the dense latewood, tangential and shear strain remain constant except for positions around the edges of the sample. A simple FE three phase growth ring model is in good agreement with the experimental values. The selective activation of individual phases like earlywood, transitionwood and latewood demonstrates that the radial hygro-expansion is dominated by the earlywood deformation, whereas tangential deformation is a complex interplay of expansion and compression that needs all tissues to fully develop.

## 1. Introduction

Hygroexpansion of Norway spruce (*Picea abies* (L.) Karst.) has been studied for long time. The well-known anisotropy with respect to the principal directions of wood (radial (R), tangential (T) and longitudinal (L)) is also well established for both hygroexpansion (e.g. Kollmann and Côté 1968) and mechanical properties (Keunecke et al. 2008; Modén and Berglund 2008b; Modén and Berglund 2008a). The wood substance strives for equilibrium with the ambient relative humidity (RH) while its porous structure significantly increases sorption velocity. Changes in the ambient RH lead to changes in the amount of water within wood, referred to as moisture content (MC). It denotes the ratio of mass of moisture over dry mass of wood. Mainly two states can be distinguished: free water (liquid or vapor) in the cell lumen and bound water within the cell wall substance (Zelinka et al. 2012). The bound water in the cell wall forms hydrogen bonds to the more or less hydrophilic wood components cellulose, hemicelluloses and lignin and drives the components apart (Simpson 1980). This, in turn, leads to dimensional changes by increase in the cell wall thickness whereas almost no in-plane swelling is observed (Ishimaru and Iida 2001). The swelling anisotropy with respect to the principal wood directions clearly originates from this behavior (Boutelje 1962; Skaar 1988). Several underlying mechanisms have been proposed to explain this anisotropy, including microfibril angle (MFA), cell arrangement, ray tissue and the alternation of earlywood (EW) and latewood (LW) bands (Frey-Wyssling 1940, 1943; Boutelje 1962; Futo 1984; Bodig and Jane 1993; Lichtenegger et al. 1999; Gindl et al. 2004; Rafsanjani et al. 2012a). Due to the effect of surrounding tissue in larger specimens or structural members, compressive and tension stresses arise (Toratti and Svensson 2000; Jönsson and Svensson 2004; Gereke et al. 2009).

On a growth ring scale, several studies show considerably different behavior for EW and LW. The properties of the cells in softwoods, mainly their wall thicknesses, differ significantly from thin walled EW cells with large lumen, developed in the beginning of the vegetation period over transitionwood (TW) to thick walled LW cells with small lumen formed towards the end of the vegetation period (e.g. (Lanvermann et al. 2013). Hence, physical properties (Young's Modulus of elasticity, density, strength) of EW considerably differ from that of TW and LW (Kretschmann and Cramer 2007; Eder et al. 2009). In the past, several investigations revealed the different hygroexpansion of EW and LW using several techniques, showing a strong anisotropy for EW while the LW showed a fairly isotropic behavior. A common method thereby was to analyze the swelling and shrinkage of isolated EW and LW samples or thin sections (Boutelje 1962; Futo 1984; Futo and Bosshard 1986; Perré and Huber 2007; Derome et al. 2011; Rafsanjani et al. 2012b).

The aim of the present work is to study perpendicular-to-grain moisture-induced deformations on a growth ring level. For this, a commercial digital image correlation solution is applied to generate full-field deformation fields. In a next step the datasets are further processed to segment the individual growth rings as well as to transform the deformations into the orthotropic radial-tangential coordinates in order to study the intra-growth ring deformations. The interpretation of the hygric behavior in terms of stress and strainfields remains incomplete, in particular the respective roles of early- and latewood on the overall behavior, since a direct measurement of stress fields is impossible. This information is obtained from accompanying Finite Element (FE) calculations.

## 2. Material and Methods

### 2.1. Test procedure and image acquisition

From a board of an approximately 107 year-old Norway spruce stem, a number of 10 samples with 40 x 40 x 5 mm$^3$ (R x T x L) were cut at positions distributed over the whole cross-section. The samples have the growth rings aligned perpendicular to one of the sample's edges. A random black and white pattern was then applied on the sample's cross-section with an airbrush gun (nozzle size 0.2mm) which is a prerequisite for digital image correlation. The samples were dried at 80°C and then equilibrated at increasing relative humidity steps (RH) (20%, 45% 65% and 95%). When they reached equilibrium, an image was taken for the subsequent evaluation of the strain fields. Therefore the optical axis of the used camera (Allied Vision Technologies, Germany), equipped with a CCD sensor with a maximum resolution of 2048 x 2048 pixel (resulting pixel size 23.5 µm), was aligned perpendicular to the radial-tangential surface. For each moisture increment the weight of the sample was recorded with a scale (precision 0.0001 g, Mettler Toledo, Switzerland).

In order to record ambient conditions throughout the experiments, a data logger (DATAQ Instruments Inc., USA) was placed next to the samples. While for the low RH of 20% the samples were stored over a saturated salt solution (lithium chloride (LiCl)), for the higher relative humidities a climate chamber (Feutron, Germany) was used. After completion of the conditioning at 95% RH, the samples were dried at 103°C, an image was taken, i.e. the reference image, and the dry masses were recorded.

### 2.2. Test evaluation procedure

The resulting images were passed to the DIC software (VIC 2D 2009, Correlated Solutions Inc., USA) which is able to track the deformation of small rectangular neighborhoods based on the principle of allocating similar gray value distributions. Herein, the oven dry image was used as the reference state. Two crucial parameters of the correlation algorithm are the *subset* and *step* parameters. The subset parameter gives the size of the rectangular neighborhood in which the cross-correlation is performed. The step parameter defines the number of pixels the center of gravity of the rectangular subset is moved until the next correlation is performed. This process is repeated until the whole area of interest is evaluated. The most accurate results, are found in our case for a subset parameter of 9 and a step parameter of 1. After the correlation, the displacement matrices were exported.

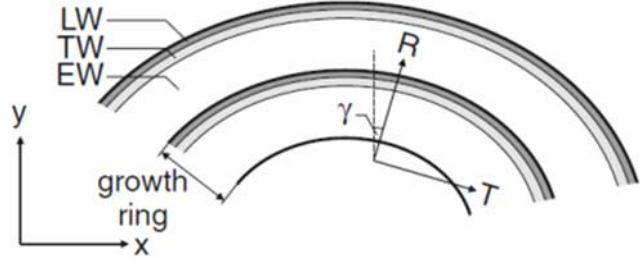

*Figure 1: Schematic growth ring.*

The further evaluation of the curved growth rings comprised the segmentation into individual rings, the interpolation of the growth rings into vectors of equal length (100 points) in the radial direction and, the averaging in the tangential direction. In order to segment the growth rings, orthographic images of the backside of the samples were acquired and the growth ring borders as well as the sample edges were drawn manually. These markings were exported as binary images that were skeletonized and transformed onto the oven dry gray scale image using the inbuilt registration tool in MATLAB®, what is basically an affine transformation with at least four reference points. With the obtained transformation function, the positions of the growth ring borders were extracted from the deformation matrix and assigned to a specific growth ring. In a next step, the vertical ($\varepsilon_{YY}$), horizontal ($\varepsilon_{XX}$) and shear strains ($\varepsilon_{XY}$), were calculated in the x/y-image coordinate system. To transform from the x/y-system to the R/T-system, the growth ring borders are approximated with spline curves. The local slope of the approximated functions is used to calculate a vector normal to the curve which corresponds to the true radial direction of the specimen. The angle $\gamma$ between the x and R axis is $dx/dy$ (see Fig. 1). The coordinate transformation then follows the following equations (local indices dropped for simplicity):

$$\varepsilon_R = \varepsilon_{XX} \sin(\gamma)^2 + \varepsilon_{YY} \cos(\gamma)^2 + \varepsilon_{XY} \sin(2\gamma)$$
$$\varepsilon_T = \varepsilon_{XX} \cos(\gamma)^2 + \varepsilon_{YY} \sin(\gamma)^2 + \varepsilon_{XY} \sin(2\gamma) \quad (1)$$
$$\varepsilon_{RT} = -\frac{1}{2}\sin(2\gamma)(\varepsilon_{XX} - \varepsilon_{YY}) + \varepsilon_{XY} \cos(2\gamma)$$

## 2.3. Three phase growth ring model

To calculate the internal residual stress distribution, a FE simulation of a block consisting of four 3mm wide growth rings that are composed of three different zones with properties for early- (EW 0.3mm), transition- (TW 0.7mm), and latewood (LW 2mm) was used. The equivalent elastic parameters and hygro-expansion coefficients from the three regions are taken from the work of Persson (2000) and summarized in *Table 1*. Note that all parameter sets fulfill consistency conditions for orthotropic materials. These properties originate from a multi-scale model and were calculated on representative cellular micro-structures of Norway spruce. A Cartesian material coordinate system is used, ignoring growth ring curvatures for simplicity. A linear thermo-elastic calculation is performed, using the analogy between moisture and temperature transport and expansion with the ABAQUS FE software. Hence, the moisture dependence of all elastic parameters and expansion coefficients, as well as all sources of material non-linearity from plastic and visco-elastic/plastic dissipation of all sorts is neglected. The FE model is shown in Fig. 8 using thermo-mechanical 20-node elements with quadratic displacement, linear temperature interpolation and reduced numerical integration (C3D20RT). Symmetry in the LR and RT plane is chosen leading to block dimensions of L-R-T of 80x12x120mm. All other 4 surfaces can displace freely. The temperature, respective moisture, is increased homogeneously for $\Delta MC = 14.2\%$. Three cases are calculated: (a) swelling of all phases (b) swelling LW and non-swelling EW/TW, and (c) EW/TW and non-swelling LW.

*Table 1: Equivalent elastic parameters and hygro-expansion coefficients for earlywood, latewood and transitionwood taken from (Persson 2000).*

| Parameter | Early wood | Transition wood | Late wood |
|---|---|---|---|
| $E_L$ [MPa] | 7710 | 11400 | 36400 |
| $E_R$ [MPa] | 671 | 953 | 1570 |
| $E_T$ [MPa] | 82.9 | 441 | 2100 |
| $G_{LR}$ [MPa] | 675 | 780 | 1760 |
| $G_{LT}$ [MPa] | 397 | 861 | 1770 |
| $G_{RT}$ [MPa] | 9.23 | 10 | 43.1 |
| $\nu_{LR}$ | 0.652 | 0.108 | 0.427 |
| $\nu_{LT}$ | 0.552 | 1.399 | 0.546 |
| $\nu_{RT}$ | 1.004 | 0.521 | 0.164 |
| $\alpha_L$ [%/%] | 0.00047 | 0.00282 | 0.00503 |
| $\alpha_R$ [%/%] | 0.23 | 0.235 | 0.335 |
| $\alpha_T$ [%/%] | 0.365 | 0.386 | 0.394 |

## 3. Results and Discussion

In the current study the hygro-expansion of wood was studied on a growth ring level on the transverse surface of wood. The full-field orthotropic strains were segmented into individual growth rings and averaged along the tangential direction (i.e. the same growth ring position). As a first step the relation between the detected strains and the distance to the sample edge is analyzed. Furthermore, the relation of the strains on a growth ring level is investigated as well as their relation with increasing MC. As a last step the moisture-dependent degree of anisotropy is discussed. Finally, the results of the FE simulations regarding the role of the different zones towards the anisotropic hygro-expansion of bulk wood is discussed.

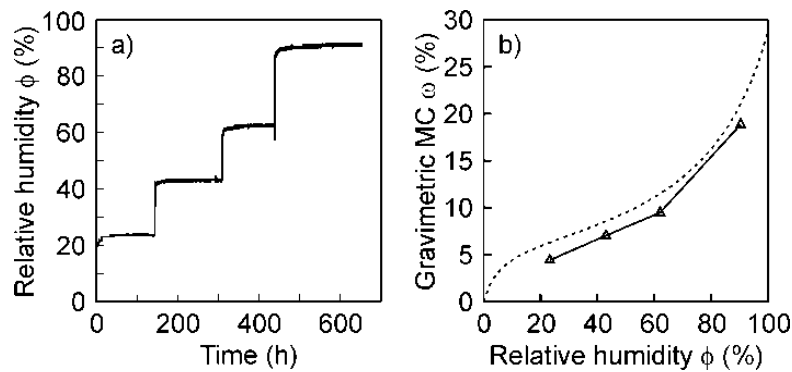

*Figure 2: Recorded relative humidity (RH) steps (a) and corresponding MC levels (b) in comparison to computed mean sorption isotherm according to the Hailwood-Horrobin model (taken from (Popper and Niemz 2009)).*

## 3.1. Sorptive behavior

The RH readings of the datalogger placed adjacent to the samples throughout the experiments are shown in

*Figure 2*a and reach from 23% up to 90% RH. The resulting sorption isotherm as the mean of all samples in comparison to the computed mean sorption isotherm according to the Hailwood-Horrobin model (taken from (Popper and Niemz 2009)) is given in

*Figure 2*b. The comparison between the model prediction and actual moisture contents leads to a good agreement, since the model represents the mean sorption isotherm and the samples were conditioned in adsorption. According to (Pang and Herritsch 2005; Moon et al. 2010; Dvinskikh et al. 2011) the presented mean MC levels are homogeneous for the entire sample since gravimetric MC is of global nature.

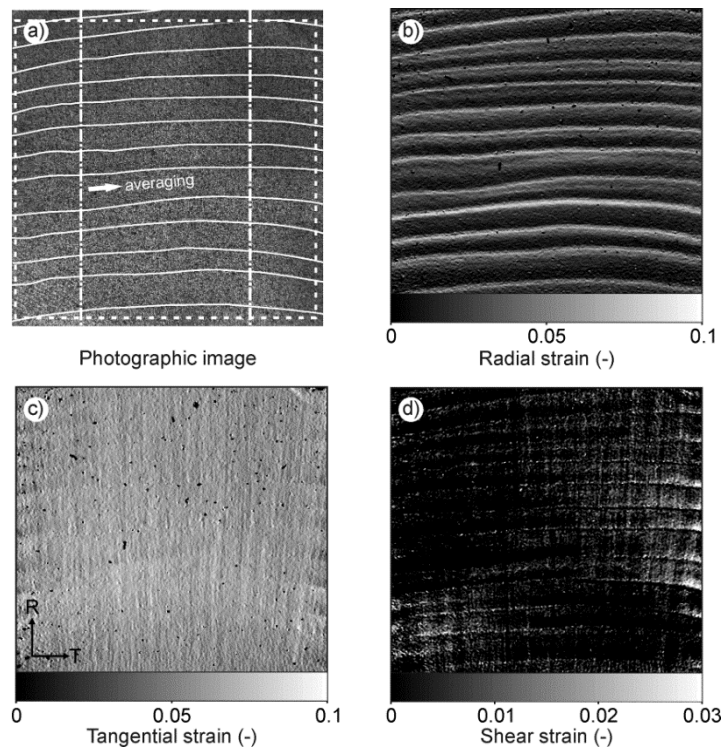

*Figure 3: Sample surface with speckle pattern (a), evaluated area of interest (AOI) (dashed line) identified growth ring boundaries (solid lines) and corresponding surface strains (b-d) at an MC of 19.2% at a RH of 90%.*

## 3.2. Full-field strain distribution

The full-field strain distributions of the evaluated area of interest (AOI) are given in Figure 3 for one sample. The $\varepsilon_{YY}$ strain field (Fig. 3c) shows pronounced differences that correspond to the growth rings, with the maxima at the growth ring borders and minima in between. In the $\varepsilon_{XX}$ strain field (Fig. 3d) no such pattern can be found, instead the strain is fairly constant within a range of 1-2%, only a slight tendency towards slightly higher values in the lower center portion can be seen. The shear strain $\varepsilon_{XY}$ (Fig. 3b) shows a slight correlation with the growth ring structure with elevated values at the growth ring borders and lower values in between. A slight concentration of higher strains can be seen in the lower left and center right position of the AOI.

However, the nature of these full-field representations is rather qualitative than quantitative. Therefore, the datasets were segmented along the growth ring borders, interpolated to 100 points in radial direction and averaged along the tangential direction, i.e. only averaging points in the same growth ring position.

### 3.3. Identification of boundary effects

In order to study the effect of boundary conditions, the growth ring wise strains are further segmented according to the regions A, B, and C (c.f. Fig. 3a) and shown in Fig. 4. The first and last columns show the mean strains of regions A and C that correspond approximately 3 growth ring widths (mean growth ring width: 2.6mm) and represent the boundary affected regions. The remaining center (region B Fig 3a) data is shown in the center column. The oscillatory behavior of the radial strain with maxima of about 5 – 7% and minima of around 1% as seen in the full-field data is clearly reflected in the segmented data. A clear influence of the position on the AOI cannot be found for $\varepsilon_R$. The tangential strain is constant at around 6% strain in the center of the AOI, whereas the outer regions show a slight wave-like trend with a variation of about 1% where the maxima correspond to the maxima in the radial strain. A similar behavior can be found for the shear strain. Whereas the outer regions show a much higher variation (about 2%), with reversed sign for the opposite sides which is a direct consequence of free boundaries, the center shows a variation around zero with much lower amplitude (around 1%), which basically can be regarded as zero.

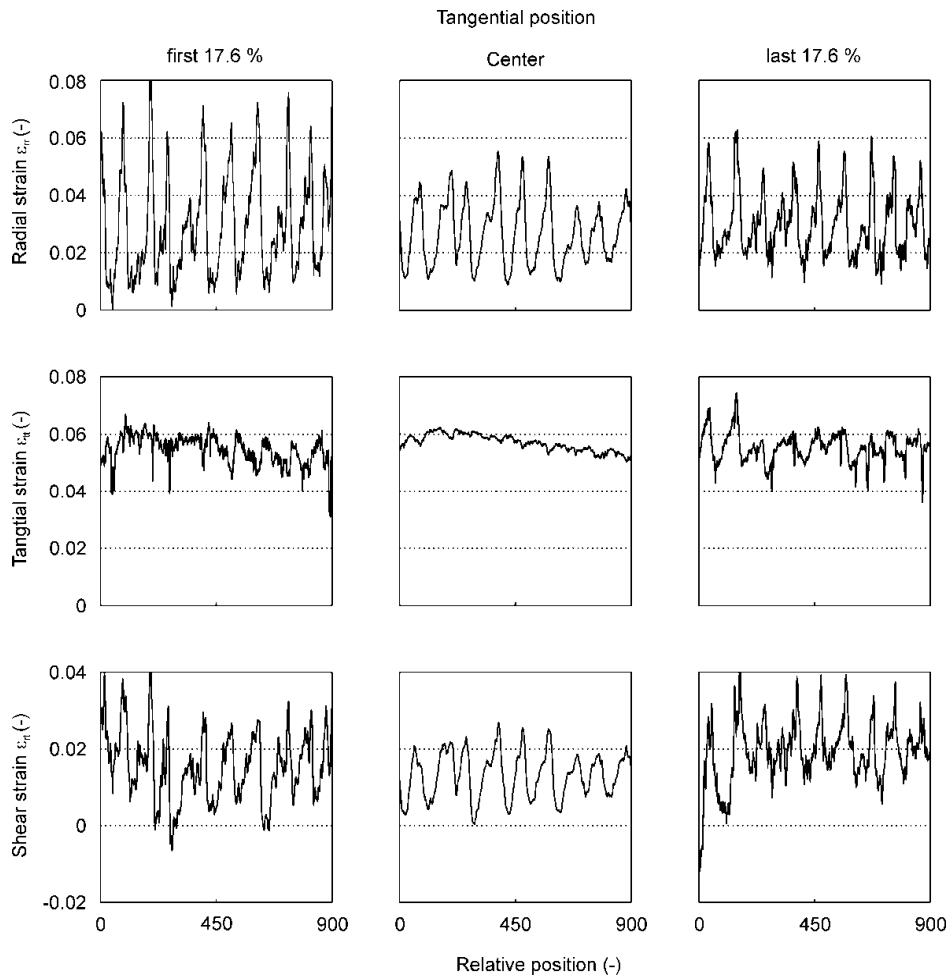

*Figure 4: Influence of sample position on detected radial, tangential, and shear strains at 19.2% MC. The first and last columns represent the boundary-affected regions (approximately 3 growth ring width).*

### 3.4. Intra-ring strain distribution

A close-up view on the hygro-expansion strains and their interrelation in one single growth ring taken from the center region of the AOI, as well as the corresponding density and MFA distribution is given in

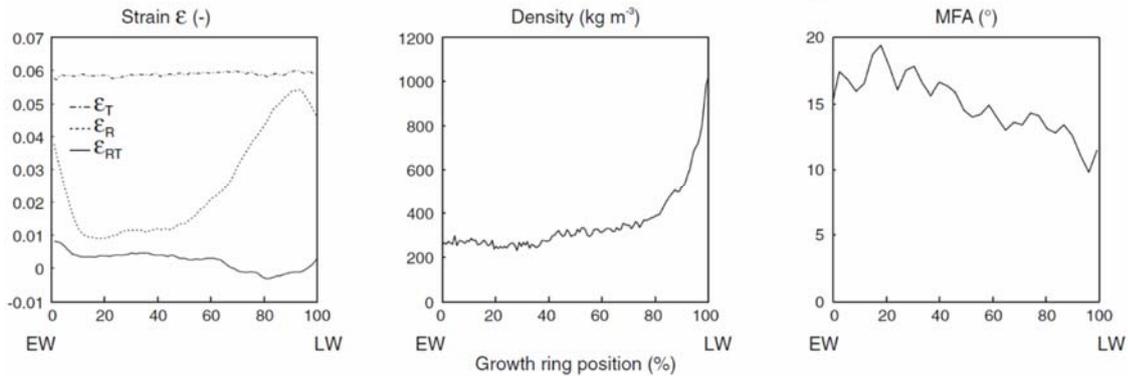

*Figure 5.* The density and MFA measurements were performed using X-ray densitometry and X-ray diffraction on samples adjacent to the present samples. For detailed information see (Lanvermann et al. 2013). Where the tangential strain ($\varepsilon_T$) can be regarded as constant as a consequence of boundary constraints, the radial strain ($\varepsilon_R$) closely follows the density progression (Spearman's $\rho$: 0.934), where the lowest strains and densities in the EW and the highest strains and densities in LW, respectively, which is in line with former measurements (Kifetew et al. 1997; Keunecke et al. 2012). The opposite relation can be found regarding $\varepsilon_R$ and MFA (Spearman's $\rho$: -0.934). The shear strain ($\varepsilon_{RT}$) varies within ±0.5% with slight positive values in EW and negative ones in LW.

### 3.5. Evolving strain distribution with increasing RH

The strains as the mean of all segmented growth rings (n=99) with increasing RH from 23% to 90% within the center of the samples are given in Figure 6. With increasing RH, and thus increasing MC, the $\varepsilon_R$ profile becomes more pronounced. This trend is clearly reflected in the radial differential swelling coefficients $\alpha_R$ for EW and LW as given in Table 2. With increasing RH $\varepsilon_T$ remains independent on the position inside the growth ring. Over all, compared to the other strains, $\varepsilon_{RT}$ is one order of magnitude smaller. Apart from the highest RH step, where a variation from -0.1% to +0.1% was found, no clear trend can be seen with increasing RH and be regarded as within the measurement uncertainty. Furthermore, regions close to the growth ring borders are prone to artifacts since the employed cross-correlation method inevitably causes a certain smoothing that can be minimized but never completely omitted.

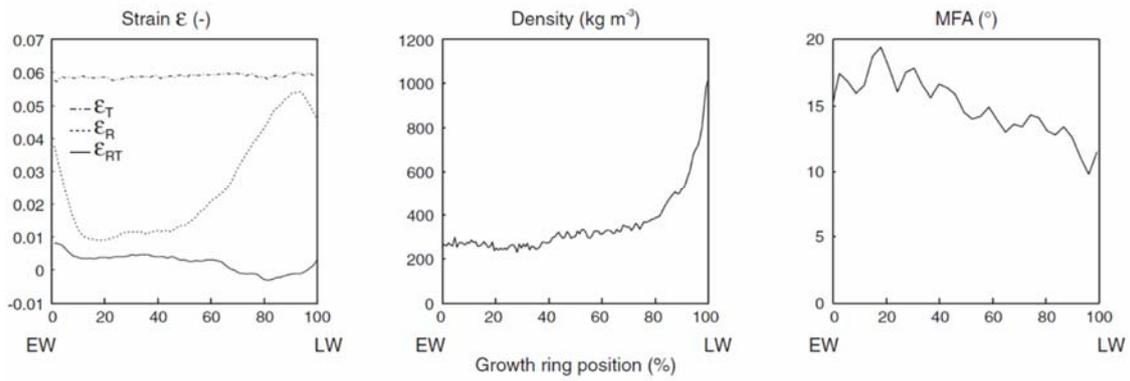

*Figure 5: Radial, tangential, and shear strain development for an individual growth ring at 19.0% MC at the center region and the corresponding density and MFA development (for detailed information see (Lanvermann et al. 2013)).*

### 3.6. Swelling Anisotropy

The anisotropic swelling (and mechanical) behavior of bulk wood is also represented on a growth ring level. The swelling ratio $\varepsilon_T / \varepsilon_R$ for the observed RH steps along a single growth ring is given in Figure 7 and shows an impressive data collapse for the different RH levels. Here, in line with other investigations (Derome et al. 2011; Rafsanjani et al. 2012b), EW shows the highest degree of anisotropy (mean: 3.90) and LW is almost isotropic (mean: 1.32) with an almost linear transition within the growth ring. Due to the constancy of $\varepsilon_T$, the anisotropic hygric behavior is solely determined by differences in $\varepsilon_R$ which leads to an overall mean swelling ratio of 2.50.

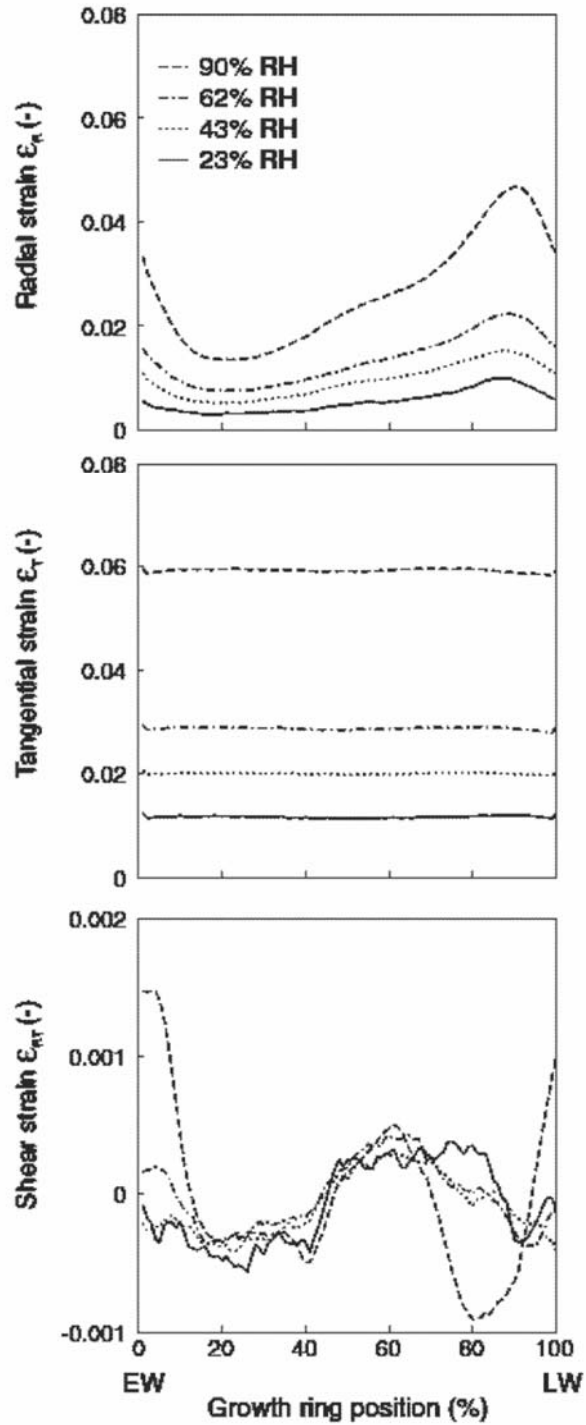

*Figure 6: Evolution of radial, tangential, and shear strains for the individual RH steps calculated as the mean of all growth rings.*

The presented method is not, however, able to resolve the underlying mechanisms that lead to the different degrees of anisotropy of EW and LW. A possible explanation regarding the different behavior of the tissues may be found in inhomogeneities in the cellular structure. But, due to the modeling approach, we are able to evaluate the contribution of EW and LW to the mean strain, which corresponds to the swelling of bulk wood.

*Table 2: Effective swelling properties (density in kg m$^{-3}$, swelling coefficients $\alpha$ in % strain/% moisture content).*

|  | Earlywood | Latewood | Bulk wood |
| --- | --- | --- | --- |

|  | $\rho_0$ | $\alpha_R$ | $\alpha_T$ | $\alpha_T/\alpha_R$ | $\alpha_R$ | $\alpha_T$ | $\alpha_T/\alpha_R$ | $\alpha_R$ | $\alpha_T$ | $\alpha_T/\alpha_R$ |
|---|---|---|---|---|---|---|---|---|---|---|
| Present work | 365 | 0.09 | 0.33 | 3.9 | 0.21 | 0.33 | 0.32 | 0.14 | 0.33 | 2.36 |
| Rafsanjani et al. 2012b. | 443 | - | - | 3.0 | - | - | 1.17 | 0.17 | 0.31 | 1.82 |
| Nakato 1958[a] | - | 0.45 | 0.45 | 1.0 | 0.50 | 0.60 | 0.2 | - | - | - |
| Persson 2000 | 400 | 0.23 | 0.36 | 1.59 | 0.33 | 0.39 | 1.18 | - | - | - |
| Kollmann and Côté 1968 | 375 | - | - | - | - | - | - | 0.19 | 0.37 | 1.95 |

[a] pure cell wall material

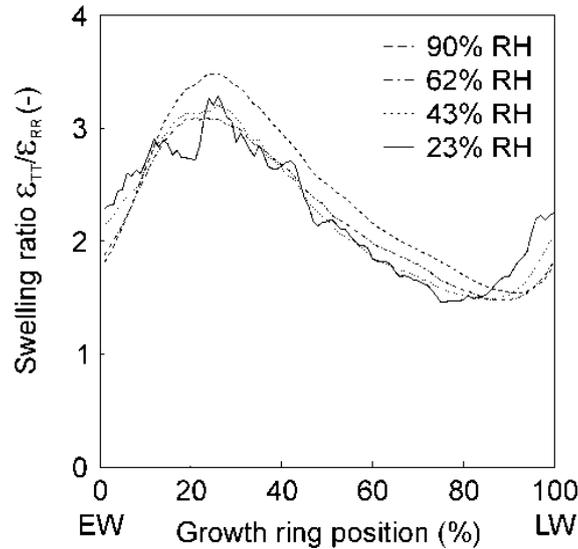

*Figure 7: Swelling ratio development within the growth rings for the individual RH steps.*

### 3.7. FE simulation

First, strains are measured along path 1 (see Fig. 8) that is along the intersection line of the two symmetry planes. The results of the simulation confronted with the measurements (Fig. 9), shows surprisingly good agreement, considering the simple background of the FE model although $\varepsilon_R$ is overestimated for EW. This overestimation clearly originates from the much higher $\alpha_R$ as found by Persson (2000) compared to the present work. For consistency reasons, it was decided to adopt the value of Persson. The steps in the radial strain for the different tissues EW, TW and LW support the observation of the strong correlation between measured strain and density inside a growth ring (Fig. 5). The effect of free surfaces is shown in Figure 11. As can be seen, RT-shear stress becomes insignificant after a distance of approximately 3 growth rings which corresponds to RT-shear strain (Fig. 3). The shear in the LR and LT does reach much less into the bulk. As result, a typical wavy surface is observed (see Fig. 11).

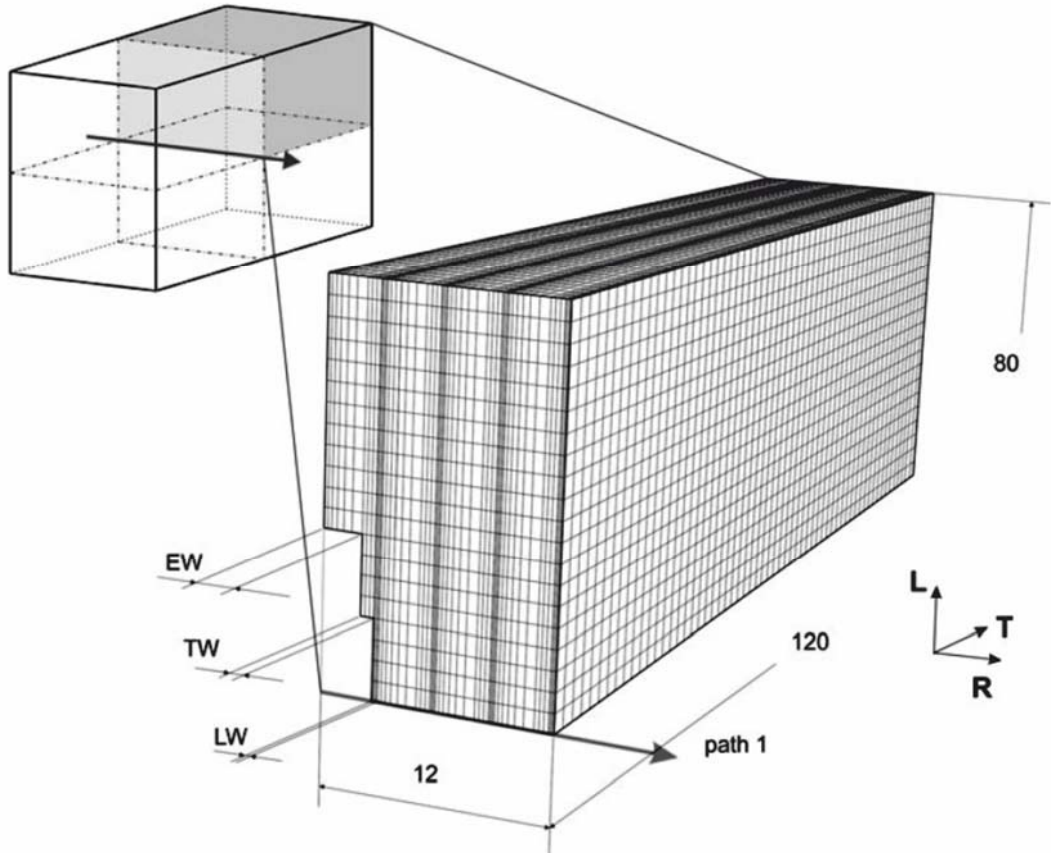

*Figure 8: FE model consisting of 4 growth rings and respective materials earlywood (EW), transitionwood (TW), and latewood (LW). All dimensions in mm. Note that the radial axis is oriented from pith to bark.*

The FE model gives the possibility not only to compute the swelling of all tissues (case (a)), but also to study the effect of isolated swelling of LW only (case (b)) or EW/TW only (case (c)). The average radial swelling strains for the three cases measured inside the system are (a) 3.39%, (b) 0.064%, and (c) 3.3%, while the local strains along path 1 (see Fig. 8) are given in Figure 10. It is very interesting to note that the radial strain basically is dominated by the EW/TW. However, the radial strain consists not only of expansion, but also of lateral contraction. For case (b), due to its expansion, LW is put under tension and compresses the EW/TW region. This results in negative swelling strains for the EW/TW regions and a very low mean strain. Vice versa is important for case (c) which leads to a radial mean strain for EW/TW swelling only that is close to case (a) where all tissues contribute to the radial swelling (see Fig. 10). The tangential strain needs swelling of all growth ring regions to fully develop. Otherwise the balance of tensile and compressive stresses leads to a decreased strain. Note that the slope of the tangential strains is due to small system deflection around the longitudinal axis caused by the limited number of growth rings in this model.

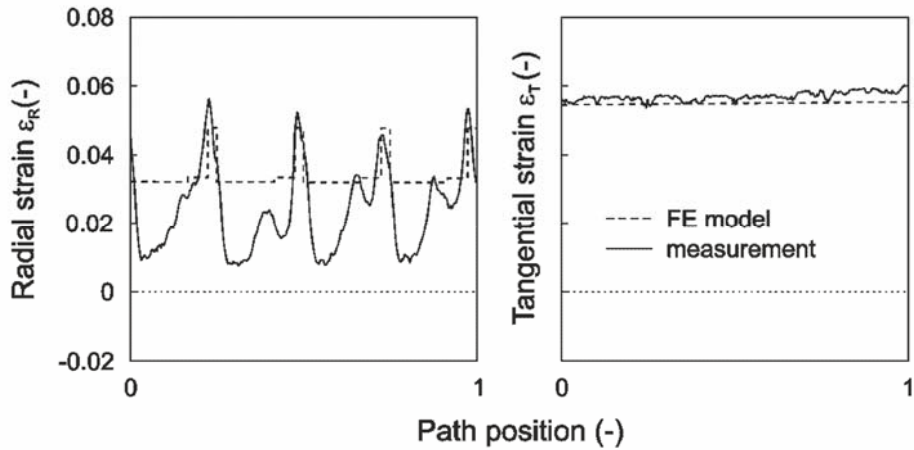

*Figure 9: FE model and measurement confrontation at 14.2% MC.*

## 4. Conclusions

In the current work, the moisture-induced swelling perpendicular-to-grain within growth rings is investigated using digital image correlation and the contribution of earlywood and latewood is modeled using a simple FEM simulation. The growth ring segmentation and averaging along local anatomic directions leads to a significant increase in quality of the data, results in a representative dataset due to the number of growth rings investigated.

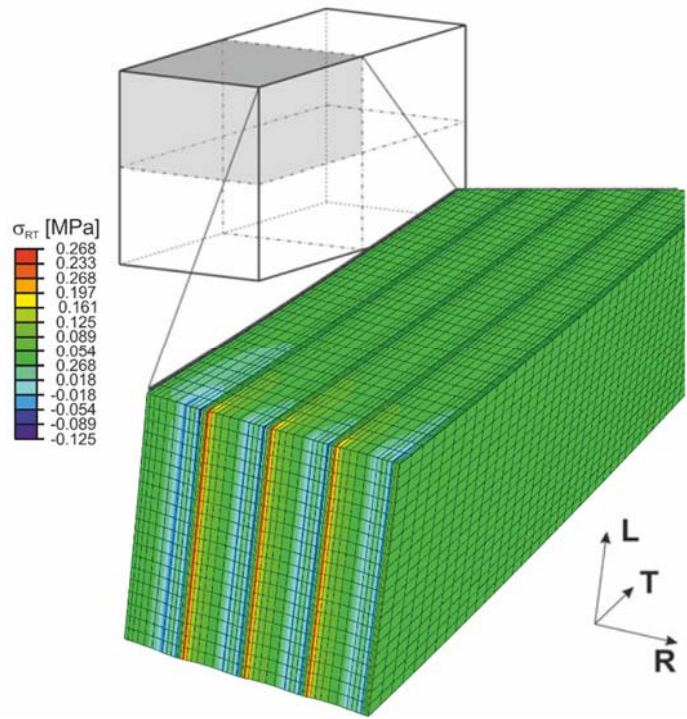

*Figure 10: Shear stress distribution at the system edges. Note that displacements are magnified by a factor of 20.*

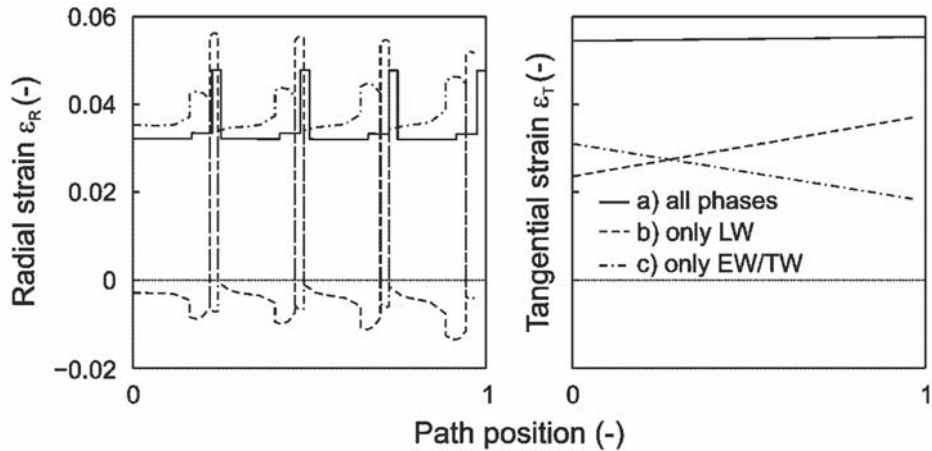

*Figure 11: Radial and tangential swelling strains along path 1 for the 3 studied cases ((a) all materials, (b) LW only and (c) LW/TW only).*

The continuous full-field principal and the growth-ring wise segmentation show a clear dependence of sample position on tangential and shear strain. Whereas tangential strain shows a pronounced correlation with high density regions resulting in a typical wavy surface, this phenomenon vanishes approximately 3 mean growth ring widths from the edges leading to a constant deformation. The same was found for the shear strain that basically is zero in the edge-unaffected center region. The close correlation between density and radial strain is clearly confirmed where the high density latewood exhibits larger strains than the low density earlywood. These relations are confirmed for all investigated relative humidity steps. The anisotropic deformation is highest in the thin-walled earlywood and gradually decreases to thick-walled latewood that is almost isotropic.

These findings are in good agreement with the simulation of a simple FEM model consisting of the three main regions, earlywood, transitionwood and latewood. Furthermore, the model reveals that the anisotropic behavior of macroscopic wood is a complex interplay of the alternation of the three wood tissues.

A potential practical application of the gained insight may be found in the design of wood bondings and coatings. Therefore, changes in the MC lead to stresses at the interface not only on a global but also on a microscale in earlywood and latewood. In future investigations, this concept should be extended to hardwoods and hygromechanical behavior on the growth ring scale.

## Acknowledgements

The authors are grateful supported by the Swiss National Science Foundation (Grant No. 125184).